\documentclass[10pt,conference,letter]{IEEEtran}
\usepackage[ansinew]{inputenc}
\usepackage[T1]{fontenc}
\usepackage{color,amsmath,enumerate,amssymb,hhline,verbatim}
\usepackage{url}
\usepackage{multirow}

\newcommand{\com}{\hbox{, }}
\newcommand{\hand}{\hbox{ and }}

\newcommand{\gpar}{(n,k,d)}
\newcommand{\lpar}{(r,\delta,t)}
\newcommand{\slpar}{(r,\delta,t)'}
\newcommand{\apart}{(n,k,d,r,\delta)}
\newcommand{\apar}{(n,k,d,r,\delta,t)}
\newcommand{\sapar}{(n,k,d,r,\delta,t)'}

\newcommand{\ncode}{{(n,k)\hbox{-code}}}
\newcommand{\ncodes}{{(n,k)\hbox{-codes}}}
\newcommand{\dcode}{{(n,k,d)\hbox{-code}}}

\newcommand{\apoly}{{\apar \hbox{-polymatroid}}}

\newcommand{\rec}{(r,\delta)}
\newcommand{\srec}{\rec'}
\newcommand{\rect}{(r,\delta,t)}
\newcommand{\srect}{(r,\delta,t)'}
\newcommand{\local}{\lpar \hbox{-availability}}
\newcommand{\slocal}{\slpar \hbox{-availability}}

\newcommand{\alp}{A}
\newcommand{\words}{\alp^n}
\newcommand{\pr}{{\mathrm{Pr}}}

\newcommand{\poly}{P = (\rho,E)}
\newcommand{\delP}{P_{|Y}}
\newcommand{\delrho}{\rho_{|Y}}
\newcommand{\delpoly}{\delP = (\delrho,Y)}
\newcommand{\cpoly}{P_C = (\rho_C,[n])}
\newcommand{\rpoly}{1_{\leq} \hbox{-polymatroid}}
\newcommand{\rpolys}{1_{\leq} \hbox{-polymatroids}}

\newcommand{\cset}{\mathcal{U}}
\newcommand{\aflat}{\mathcal{F}}
\newcommand{\cflat}{\mathcal{Z}}
\newcommand{\botcflat}{0_\cflat}
\newcommand{\topcflat}{1_\cflat}
\newcommand{\cl}{\mathrm{cl}}

\begin{document}

\title{Applications of Polymatroid Theory to Distributed Storage Systems}


\author{
\authorblockN{Thomas Westerb{\"a}ck$^*$}
\authorblockA{$^*$Department of Mathematics and Systems Analysis\\ Aalto University, P.O. Box 11100\\ FI-00076 Aalto, Finland \\ (e-mails: firstname.lastname@aalto.fi)}
\and
\authorblockN{Ragnar Freij-Hollanti$^\dag$ and Camilla Hollanti$^*$}
\authorblockA{$^\dag$Department of Communications and Networking\\ Aalto University, P.O. Box 13000\\ FI-00076 Aalto, Finland\\  (e-mail: firstname.lastname@aalto.fi)}
}

\maketitle

\newtheorem{definition}{Definition}[section]
\newtheorem{thm}{Theorem}[section]
\newtheorem{proposition}[thm]{Proposition}
\newtheorem{lemma}[thm]{Lemma}
\newtheorem{corollary}[thm]{Corollary}
\newtheorem{exam}{Example}[section]
\newtheorem{conj}{Conjecture}
\newtheorem{remark}{Remark}[section]
\newtheorem{set_construction}{A construction of matroids}[section]

\begin{abstract}

In this paper, a link between polymatroid theory and locally repairable codes (LRCs) is established. The codes considered here are completely general in that they are subsets of $\alp^n$, where $\alp$ is an arbitrary finite set. Three classes of LRCs are considered, both with and without availability, and for both information-symbol and all-symbol locality. The parameters and classes of LRCs are generalized to polymatroids, and a generalized Singelton bound on the parameters for these three classes of polymatroids and LRCs is given. This result generalizes the earlier Singleton-type bounds given for LRCs. Codes achieving these bounds are coined \emph{perfect}, as opposed to the more common term \emph{optimal} used earlier, since they might not always exist. Finally, new constructions of perfect linear LRCs are derived from gammoids, which are a special class of matroids. Matroids, for their part, form a subclass of polymatroids and have proven useful in analyzing and constructing linear LRCs.
\end{abstract}

\section{Introduction}

Within the past few years, \emph{distributed storage systems} (DSSs) have revolutionized our traditional ways of storing, securing, and accessing data, and various big players like Facebook and Google nowadays provide their own cloud storage services. However, they do not come without regular failures, and hence have to be maintained by sophisticated repair processes. It has turned out that the number of nodes contacted for repair forms a bottle-neck in such vast data centers, calling for the notion of \emph{locality}. In addition, clusters containing hot data, namely data that is frequently accessed simultaneously by  many users, will benefit from multiple repair alternatives. This feature has further motivated the notion of \emph{availability}.

\subsection{Locally Repairable Codes}

In this paper, we consider \emph{locally repairable codes (LRCs)} with availability from the viewpoint of the interplay between its global parameters $(n,k,d)$ and local parameters $(r,\delta, t)$. We will consider $\apart$-LRCs, $\apar$-LRCs, and $\sapar$-LRCs with 1-information-symbol locality, information-symbol locality, and/or all-symbol locality. These parameters and notions will be explained in detail in the sequel. 

Let $\alp$ be a finite set of size $s$ and $C$ a nonempty subset of $\words$. Then we call $C$ an $\ncode$, where $k = \log_s(|C|)$. For $X = \{x_1,\ldots,x_l\} \subseteq [n] = \{1,\ldots,n\}$ and $\boldsymbol{z} \in \words$, let $\boldsymbol{z}_X = (z_{x_1},\ldots,z_{x_l})$. The \emph{projection} of $C$ into $\alp^{|X|}$ is defined as 
$$
C_X = \{\boldsymbol{c}_X = (c_{x_1},\ldots,c_{x_l}): \boldsymbol{c} \in C\,, |X|=l\}.
$$

The \emph{minimum (Hamming) distance} $d$ of $C$ can be defined as
\begin{equation} \label{eq:hamming}
d = \min \{|X|: X \subseteq [n] \hbox{ and } |C_{[n] \setminus X}| < |C|\}.
\end{equation}
In other words, for any $X\subseteq [n]$ with $|X|<d$, the symbols in $X$ can be reconstructed by observing the symbols in $[n]\setminus X$ for every codeword in $C$, whereas for $|X|=d$ this is not necessarily true anymore.

We will consider two types of repair sets, $\rec$ and $\srec$, where $r,\delta \in \mathbb{Z}$, $r \geq 1$, and $\delta \geq 2$. To this end, let $i \in [n]$ be a code symbol, equivalently a storage node, and $R\subseteq [n]$. The set $R \subseteq [n]$ is a \emph{local repair set} with \emph{repair locality} $\rec$ for the node $i$ if  
$$
\begin{array}{rl}
(i) & i \in R\,,\\
(ii) & |R| \leq r+\delta-1\,,\\
(iii) & X \subseteq R \setminus \{i\} \,\com |X| = |R| - (\delta - 1) \Rightarrow |C_{X}| = |C_R|\,.
\end{array}
$$
The set $R \subseteq [n]$ is a \emph{local repair set} with \emph{repair locality} $\srec$ for the node $i$ if the conditions $(i)$ and $(ii)$ above are satisfied, and in addition we have that
$$
\begin{array}{rl}
(iii)' & X \subseteq R \,\com |X| = |R| - (\delta - 1) \Rightarrow |C_{X}| = |C_R|\,.
\end{array}
$$

For $\rec$-locality, the condition $(iii)$ means that, for all codewords and subsets $X \subseteq R \setminus \{i\}$ such that $|X| \geq |R| - (\delta - 1)$, the symbols indexed by $X$ are always sufficient to recover the symbol indexed by $i$. Also, the minimum distance of $C_{R \setminus \{i\}}$ is equal to or greater than $\delta-1$.

For $\srec$-locality, the condition $(iii)'$ means that, for all codewords and subsets $X \subseteq R \setminus \{i\}$ such that $|X| \geq |R| - (\delta - 1)$, the symbols indexed by $X$ are always sufficient to recover the symbol indexed by $i$. Also, the minimum distance of $C_{R}$ is equal to or greater than $\delta$.

Further, a coordinate $i \in [n]$ has \emph{$\local$} (resp. \emph{$\slocal$}) if there are $t$ local repair sets $R_1,\ldots,R_t$  for $i$ with $\rec$-locality (resp. $\srec$-locality) such that 
$$
\begin{array}{rl}
(iv) & j \neq l \Rightarrow R_j \cap R_l = \{i\}\,. 
\end{array}
$$
A subset $X \subseteq [n]$ has $\local$ (resp. $\slocal$) if all elements $i \in X$ have $\local$ (resp. $\slocal$).

A subset $K  \subseteq [n]$ such that 
$$
|C_K| = |C| \hand |C_{K \setminus \{i\}}| < |C|
$$
for each element $i \in K$ is called an \emph{information set}. This means that for any codeword the symbols indexed by $K$ are enough to reconstruct all the other symbols of the codeword, but a strict subsets of these symbols is not.

Moreover, a \emph{1-information set} $K$ is an information set with the additional property that for every coordinate $i \in K$ and symbols $a,b \in C_{\{i\}}$,
$$
|\{\boldsymbol{c} \in C : c_i = a \}| = |\{\boldsymbol{c} \in C : c_i = b \}|.
$$
For example, a \emph{systematic}  $\ncode$ is a code for which $k$ is an integer and there is an information set $K$ of size $k$. This yields that $K$ is a 1-information set where 
$$
|\{\boldsymbol{c} \in C : c_i = a \}| = |\alp|^{k-1},
$$ 
for each $i \in K$ and symbol $a \in \alp$.

Let $C$ be an $\dcode$ and $X \subseteq [n]$. Then $C$ is an $\apart$-LRC, $\apar$-LRC, or $\sapar$-LRC over $X$ if all elements in $X$ have $(r,\delta,t=1)$-locality, $\lpar$-availability, or $\slpar$-availability, respectively. If $X$ is an information set, 1-information set, or $X = [n]$, then  $C$ has \emph{information-symbol locality}, \emph{1-information-symbol locality}, or \emph{all-symbol locality}, respectively. 

By a \emph{linear} $(n,k)$-LRC we mean a subspace $C$ of dimension $k$ of $\mathbb{F}_q^n$, where $\mathbb{F}_q$ denotes the finite field of size $q$. 

\subsection{Related Work}

There are several papers on different Singleton-type bounds for scalar, vector-linear, and nonlinear LRCs over finite fields,  \cite{singleton64, gopalan12, prakash12, papailiopoulos12, silberstein13, rawat14} among others. Using entropy to analyze LRCs has, for example, been used in \cite{papailiopoulos12, rawat14, wang14}. Further, combinatorial methods have been used for LRCs, \emph{e.g.}, by the concept of regenerating sets \cite{wang14} and matroids \cite{tamo13, westerback15}. For LRCs with availability, some constructions are proposed in \cite{wang14b, rawat14, wang15}.

\subsection{Contributions and Organization}

Every linear code has an associated matroid, however codes in general cannot be associated with a matroid. In this paper we extend the work in \cite{westerback15} on how matroids and linear LRCs are connected by associating any LRC over any finite set $\alp$ with a polymatroid. Matroids are a subclass of polymatroids. Especially, we prove that the parameters associated with an LRC can be determined by its associated polymatroid. Moreover, we generalize the parameters associated with LRCs to polymatroids. Then, by using polymatroid theory, we get Singleton-type bounds for polymatroids which generalizes the bounds given in  \cite{singleton64, gopalan12, prakash12, papailiopoulos12, silberstein13, rawat14} for LRCs. Moreover, these new bounds also give novel bounds on LRCs, since we simultaneously consider the parameters $\delta$ and $t$, alphabets that are not finite fields or finite vector spaces, and codes of size that is not a power of a prime. 

A construction of linear LRCs with availablity is given, making use of an earlier result on a construction of linear LRCs from matroid theory in \cite{westerback15}. By this construction we are able to obtain a class of perfect linear LRCs with availability including all the parameters $\lpar$. All the parameters for a class of perfect linear LRCs considering the availability $(r,\delta,t=2)$ given in \cite{wang14b} are included in our construction.

In Section \ref{sec:polymatroids-codes}, we give some fundamentals on polymatroid theory and entropy, and describe how codes $C \subseteq \alp^n$ can be associated to a polymatroids by the use of entropy. In Section \ref{sec:LRC-polymatroid}, the associated parameters of an LRC are generalized to polymatroids and bounds on these parameters for polymatroids and LRCs are given. In Section \ref{sec:construction}, a construction of linear LRCs is given which we then use to get a class of perfect LRCs with $\lpar$-availability.

\section{Polymatroids and Codes} \label{sec:polymatroids-codes}

In this section we will show how $\ncodes$ can be associated to polymatroids via the notion of entropy. For more information on polymatroids, we refer the reader to \cite{oxley11}.

\subsection{Overview of Polymatroid Theory}

For a finite set $E$, let $2^E$ denote the collection of all subsets of $E$. A pair $P = (\rho,E)$ is a (finite) \emph{polymatroid} on $E$ with a \emph{set function} $\rho: 2^E \rightarrow \mathbb{R}$ if $\rho$ satisfies the following three conditions for all subsets $X,Y \subseteq E$:
$$
\begin{array}{rl}
(R1) & \rho(\emptyset) = 0\,,\\
(R2) & X \subseteq Y \Rightarrow  \rho(X) \leq \rho(Y)\,,\\
(R3) & \rho(X) + \rho(Y) \geq \rho(X \cup Y) + \rho(X \cap Y)\,.
\end{array}
$$
A \emph{matroid} is a polymatroid which additionally satisfies the following two conditions for all $X \subseteq E$:
$$
\begin{array}{rl}
(R4) & \rho(X) \in \mathbb{Z}\,,\\
(R5) & \rho(X) \leq |X|\,.
\end{array}
$$

For any polymatroid $\poly$ and $Y \subseteq E$ we obtain a new polymatroid $\delpoly$, where 
$$
\delrho(X) = \rho(X)
$$
for any $X \subseteq Y$.

A polymatroid $\poly$ for which $\rho(\{x\}) \leq 1$ for all $x \in E$ is called a \emph{$\rpoly$} throughout the paper. We say that two polymatroids $\poly$ and $P' = (\rho',E)$ on the same ground set $E$ are equivalent if there is a constant $c\in \mathbb{R}$ such that $\rho(X)=c\rho'(X)$ for each $X\subseteq E$. Clearly, any polymatroid $P =(\rho,E)$ is equivalent to a $\rpoly $, wherefore we will only consider $\rpolys$ for the rest of the paper. Note that if $\poly$ is a $\rpoly$, then $\delpoly$ is also a $\rpoly$ for every subset $Y \subseteq E$.

\subsection{Some Basic Properties and Notions for $\rpolys$}

The axioms (R1) and (R3) imply the following proposition.

\begin{proposition} \label{prop:rank_1_poly}
Let $\poly$ be a $\rpoly$. Then for any subset $X \subseteq E$ and element $x \in X$,
$$
\begin{array}{rl}
(i) & \rho(X) \leq |X|\,,\\
(ii) & 0 \leq \rho(X) - \rho(X \setminus \{x\}) \leq 1\,.
\end{array}
$$
\end{proposition}

\begin{IEEEproof}
The statement in (i) follows by induction on $|X|$: it is trivially true for $|X|=0$; now let $y \in E \setminus X$. Then by the induction assumption and the axioms (R1) and (R3),
$$
\rho(X \cup \{y\}) \leq \rho(X) + \rho(\{y\}) \leq |X| + 1.
$$  
For statement (ii), by axiom (R2), we immediately obtain that $0 \leq \rho(X) - \rho(X \setminus \{x\})$. Further,  by the axioms (R1) and (R3),
$$
\rho(X) - \rho(X\setminus \{x\}) \leq \rho(\{x\}) \leq 1.
$$  
\end{IEEEproof}

By generalizing some notions from matroid theory we get the following corresponding notions for any $\rpoly$ $\poly$ and $X \subseteq E$,
$$
\begin{array}{cl}
(i) & \eta(X) := |X| - \rho(X),\\
(ii) & \cl(X) := \{y \in E : \rho(X \cup \{y\}) = \rho(X)\},\\
(iii) & X \hbox{ is a \emph{flat} if } \cl(X) = X,\\
(iv) & X \hbox{ is \emph{cyclic} if for all elements } x \in X, \\
      & \rho(X) - \rho(X \setminus \{x\}) < 1.
\end{array}
$$
The collection of flats, cyclic sets, and cyclic flats of $P$ are denoted by $\aflat$, $\cset$ and $\cflat$ respectively. Note that by definition $\emptyset \in \cset$.

The following proposition will be needed for proving our generalized Singleton bound later on. 

\begin{proposition} \label{prop:mixed_1_poly}
Let $\poly$ be a $\rpoly$, then for any subsets $X,Y \subseteq E$,
$$
\begin{array}{rl}
(i) & \eta(X) \leq \eta(X \cup Y),\\
(ii) & \rho(\cl(X)) = \rho(X),\\
(iii) & X \subseteq Y \Rightarrow \cl(X) \subseteq \cl(Y),\\
(iv) & X \in \aflat \com x \in X \com \rho(X \setminus \{x\}) < \rho(X) \Rightarrow \\
     & (X \setminus \{x\}) \in \aflat,\\
(v) &  X' \subseteq X \subseteq Y \Rightarrow \\
     & \rho(X) - \rho(X \setminus X') \geq \rho(Y) - \rho(Y \setminus X'),\\
(vi) & X,Y \in \cset \Rightarrow X \cup Y \in \cset,\\
(vii) & X,Y \in \cset \Rightarrow \cl(X \cup Y) \in \cflat,\\

\end{array}
$$
\end{proposition}

\begin{IEEEproof} For a proof of the results above we use some basic facts about polymatroids. A proof will appear in the journal version of this paper.

\end{IEEEproof}

\subsection{Codes and Entropy}

We can associate any $\ncode$ with a random vector $\boldsymbol{Z}=(Z_1,\ldots,Z_n)$ with a joint probability distribution by
$$
\pr(\boldsymbol{Z} = \boldsymbol{z}) = 
\left \{
\begin{array}{ccl} 
1/|C| & \hbox{if} & \boldsymbol{z} \in C,\\
0 & \hbox{if} & \boldsymbol{z} \notin C.
\end{array}
\right .
$$
This gives, for the projections of the code, that 
\begin{equation} \label{eq:pr_joint}
\pr(\boldsymbol{Z}_X = \boldsymbol{z}_X) =  
|\{\boldsymbol{c} \in C: \boldsymbol{c}_X = \boldsymbol{z}_X \} | / |C|,
\end{equation}
where $X = \{x_1,\ldots,x_l\} \subseteq [n]$, $\boldsymbol{Z}_X = (Z_{x_1},\ldots,Z_{x_l})$ and $\boldsymbol{z}_X \in \alp^{|X|}$. The \emph{joint entropy} function of $\boldsymbol{Z}_X$ is then defined by using this probability as
\begin{equation} \label{eq:H}
H_C(\boldsymbol{Z}_X) = \sum_{\boldsymbol{z}_X \in \alp^{|X|}} \pr(\boldsymbol{Z}_X = \boldsymbol{z}_X) \log_s \left (\frac{1}{\pr(\boldsymbol{Z}_X = \boldsymbol{z}_X)} \right ),
\end{equation}
where again $s=|\alp|$, and where we have the conventions that $0\log_s 0=0$ and  $H_C(\boldsymbol{Z}_\emptyset) = 0$.

\subsection{Codes and Their Representations as Polymatroids}

From \cite{fujishige78} we have the following theorem.

\begin{thm}
The joint entropy function $H$ of any random vector $(Z_1,\ldots,Z_n)$ over some underlying probability space defines a polymatroid $P = (\rho,[n])$, where for any subset $X \subseteq [n]$
$$
\rho(X) = H(\boldsymbol{Z}_X).
$$
\end{thm}

 Hence, by \eqref{eq:pr_joint} and \eqref{eq:H}, every $\ncode$ $C$ over $\words$ induces a polymatroid $\cpoly$ where
$$
\rho_C(X) = H_C(\boldsymbol{Z}_X)\,.
$$

By the above formula and the log sum inequality we obtain the following proposition. 

\begin{proposition} \label{pro:P_C}
Let $C$ be an $\ncode$ over $\alp$ with $|\alp|=s$. Then for the polymatroid $\cpoly$ and for subsets $X,Y \subseteq [n]$,
$$
\begin{array}{rl}
(i) & P_C \hbox{ is a } \rpoly,\\
(ii) & |C_{X \cup Y}| > |C_X| \iff \rho_C(X \cup Y) > \rho_C(X),\\
(iii) & |C| = s^{\rho_C([n])},\\
(iv) & |C|/|\words| = s^{\rho_C([n]) - n}.
\end{array}
$$
\end{proposition}

\begin{exam}
From this definition of a polymatroid $\cpoly$ we get the following characterization of certain classes of codes. Let again $A$ be a finite set of size $s$. 
\begin{itemize}
\item Linear codes over $\mathbb{F}_q$: $\rho_C(X) = \log_q(|C_X|) = \mathrm{rank}(\hbox{generator matrix over column-set $X$})\in\mathbb{Z}$,
\item Almost affine codes: $\rho_C(X) = \log_q(|C_X|) \in \mathbb{Z}$,
\item Vector-linear codes, \emph{i.e.}, $C$ is a linear subspace of $\alp^n$, where $\alp = \mathbb{F}_q^\alpha$): $\rho_C(X) = \log_{q^\alpha}(|C_X|)\in\mathbb{R}$,
\item Quasi-uniform codes over $\alp$: $\rho_C(X) = \log_s(|C_X|)\in\mathbb{R}$,
\item A general code  $C \subseteq A^n$: $\rho_C(X) = H_C(\boldsymbol{Z}_X)\in\mathbb{R}$.
\end{itemize}
\end{exam}

\section{LRCs and Polymatroid Theory} \label{sec:LRC-polymatroid}

\subsection{Code Parameters for Polymatroids}

A parameter of a code is \emph{polymatroid invariant} if it only depends on its associated polymatroid, \emph{i.e.}, always has the same value on two codes with the same associated polymatroid.

We claim that the parameters $\gpar$ of a code $C\subseteq \alp^n$, $\rect$-availability and $\srect$-availability of a code symbol, as well as information-set locality and 1-information-set locality, are all polymatroid invariant properties of $C$. This follows from the definitions of these properties and the set function $\rho_C$ by using projections and  Proposition \ref{pro:P_C}. Hence, we can naturally generalize the typical code parameters to $\rpolys$.

\begin{definition} \label{def:polymatroid_parameters}
Let $\poly$ be a $\rpoly$. Then
$$
\begin{array}{rl}
(i) & n = |E|,\\
(ii) & k = \rho(E),\\
(iii) & d = \min \{|X| : \rho(E \setminus X) < \rho(E)\},\\
(iv) & \hbox{if we let $x \in E$ and $r, \delta \in \mathbb{Z}$, where } r \geq 1 \hand \delta \geq 2,\\
& \hbox{then $x$ has $\local$} \hbox{ if there are $t$ subsets}\\
     & \hbox{$R_1,\ldots,R_t \subseteq E$ such that for $i,j \in [t]$:}\\
     & \begin{array}{rl}
       (a) & x \in R_i,\\
       (b) & |R_i| \leq r + \delta - 1,\\
       (c) & Y \subseteq R_i \setminus \{x\} \com |Y| = |R_i| - (\delta - 1) \Rightarrow\\
           & \rho(Y) = \rho(R_i),\\
       (d) & i \neq j \Rightarrow R_i \cap R_j = \{x\},
       \end{array}\\
& \hbox{Similarly, $x$ has $\slocal$} \hbox{ if there are $t$ subsets}\\
     & \hbox{$R_1,\ldots,R_t \subseteq E$ such that the conditions (a), (b)}\\
     & \hbox{and (d) above are satisfied, and in addition}\\
     & \begin{array}{rl}
       (e) & Y \subseteq R_i \,\com |Y| = |R_i| - (\delta - 1) \Rightarrow\\
           & \rho(Y) = \rho(R_i),\\
       \end{array}\\
(v) & \hbox{$K \subseteq E$ is an information set if $\rho(K) = k$ and} \\
    &  \rho(K \setminus \{ x\} < k) \hbox{ for all } x \in K,\\
(vi) & \hbox{$K \subseteq E$ is a 1-information set if $K$ is an}\\
    & \hbox{information set and $\rho(x) = 1$ for all $x \in K$.}
\end{array}
$$ 
\end{definition}

Let now $x \in E$ and $R \subseteq E$. If the conditions (a)-(c) above are satisfied by $x$ and $R$ then, similarly as for codes, $R$ is called a  \emph{local repair set} with \emph{repair locality} $\rec$ for $x$. Further, if the conditions (a), (b), and (e) above are satisfied by $x$ and $R$, then $R$ is again called a  \emph{local repair set} with \emph{repair locality} $\srec$ for $x$.

We remark that the values of the parameters $\gpar$, $\lpar$, and $\slpar$ for a code $C$ and a node $i$ are the same as for the associated polymatroid $P_C$ and its element $i$. Further, a coordinate set $K$ for a code $C$ is an information set (resp. 1-information set) if and only if the corresponding set of elements $K$ in $P_C$ is an information set (resp. 1-information set).

\subsection{Code Parameters in Terms of Cyclic Flats}

Let $\poly$ be a $\rpoly$. First we remark that $d$ is well-defined for any nontrivial $P$, that is, for any $P$ whose set function $\rho$ is not the zero function. Second, if there is an element $x \in E$ such that $x$ is not in any cyclic flat, then $\rho(E \setminus \{x\}) = \rho(E) - 1$. This implies that $\rho(X-\{x\}) = \rho(X) - 1$ for all  $X \subseteq E$ with $x \in X$.  This, for its part, implies that there are no repair sets for $x$. Further, let $R$ be a repair set of $y$ with repair locality $\rec$ (resp. $\srec$) and $x \in R$.  Then $R \setminus \{x\}$ is a repair set of $y$ with repair locality $(r-1, \delta)$ (resp. $(r-1,\delta)'$).  Consequently, we are only interested in $\rpolys$ $\poly$ for which $k \neq 0$ and the union of cyclic flats, denoted by $\topcflat$, is the whole set $E$.

The following proposition gives a list of basic facts that will be needed later.

\begin{proposition} \label{proposition:recovering_sets}
Let $\poly$ be a $\rpoly$. Then for any element $x \in E$ and subsets $X,Y \in \cset$,
$$
\begin{array}{rl}
(i) & if \hbox{$R$ is a repair set of $x$ with $\rect$-}\\
    & \hbox{locality, then there is a repair set}\\
    & \hbox{$Q \subseteq R$ of $x$ with $\local$},\\
(ii) & if \hbox{$R'$ is a repair set of $x$ with $\srect$-}\\
    & \hbox{locality, then there is a repair set}\\
    & \hbox{$Q' \subseteq R'$ of $x$ with $\slocal$},\\
(iii) & \cl(X) \in \cflat,\\
(iv) & \cl(X \cup Y) = \cl(\cl(X) \cup \cl(Y)) \in \cflat,\\
(v) & \rho(X) \leq |X| - (\delta-1),\\
(vi) & \eta(X) \geq \delta-1,\\
(vii) & \rho(X \cup Y) \leq \rho(X) + \rho(Y) - \rho(X \cap Y),\\
(viii) & \eta(X \cup Y) \geq \eta(X) + \eta(Y) - \eta(X \cap Y). 
\end{array}
$$
\end{proposition}

\begin{IEEEproof}
For a proof of the results above we use some basic facts about polymatroids. A proof will appear in the journal version of this paper.
\end{IEEEproof}

We are now ready to connect the parameters $(n,k,d)$, $\rect$ and $\srect$ of a polymatroid using cyclic flats.

\begin{thm} \label{thm:parameters_via_Z}
Let $\poly$ be a $\rpoly$ with $k > 0$ and $1_\cflat = E$. Then 
$$
\begin{array}{rl}
(i) & n = \lvert 1_\cflat \rvert,\\
(ii) & k = \rho(1_\cflat),\\
(iii) & d = \left \lfloor n - k + 1 - \max \{ \eta(Y): Y \in \cflat \setminus \{1_\cflat \}\} \right \rfloor,\\
(iv) & \hbox{$x \in E$ has $\local$ if and only if there }\\
     & \hbox{are $t$ repair sets $R_1,\ldots,R_t \in \cset$ with repair locality}\\
     & \hbox{$\rec$, all of whose pairwise intersections equal $\{x\}$},\\
(v) & \hbox{$x \in E$ has $\slocal$ if and only if there }\\
     & \hbox{are $t$ repair sets $R_1',\ldots,R_t' \in \cset$ with repair locality}\\
     & \hbox{$\srec$, all of whose pairwise intersections equal $\{x\}$}.
\end{array}
$$
\end{thm}

\begin{IEEEproof}
The statements (i) and (ii) follow directly from Definition \ref{def:polymatroid_parameters}. The statement (iv) follows from Proposition \ref{proposition:recovering_sets}(i) and Definition \ref{def:polymatroid_parameters}. Similarly,  statement (v) follows from Proposition \ref{proposition:recovering_sets}(ii) and Definition \ref{def:polymatroid_parameters}. 

For (iii), we first obtain that
\begin{equation} \label{eq:d}
\begin{array}{rl}
d & = n - \max \{|Y| : Y \subseteq E \hbox{, } \rho(Y) < k\}\\
  & = \max \{n - |Y| : Y \subseteq E \hbox{, } k-1 \leq \rho(Y) < k\}\\
  & = \max \{\left \lfloor n - |Y| - k+1  + \rho(Y) \right \rfloor: Y \subseteq E \hbox{, } \rho(Y) < k\}\\
  & = \left \lfloor n - k + 1 - \max \{ \eta(Y) : Y \in \cflat \setminus \{\topcflat\}\} \right \rfloor.
\end{array}
\end{equation}
In the equations above, the first equality is a consequence of \eqref{eq:hamming}, and the second of Axiom (R2) and Proposition \ref{prop:rank_1_poly}(i). Further, for $X \subseteq E$,
\begin{equation} \label{eq:eta_X_clX}
\begin{array}{rl}
(i) & \eta(X) \leq \eta(\cl(X)),\\
(ii) & x\in X \com \rho(X \setminus \{x\}) = \rho(X) -1 \Rightarrow\\
     & \eta(X \setminus \{x\}) = \eta(X).
\end{array}
\end{equation}
Inequality \eqref{eq:eta_X_clX}(i) is a consequence of Proposition \ref{prop:mixed_1_poly}. Now, by \eqref{eq:d}, \eqref{eq:eta_X_clX} and by the fact that $\cl(\emptyset)$ is a cyclic flat and a subset of all flats, we obtain that
$$
d  = \left \lfloor n - k + 1 - \max \{ \eta(Y) : Y \in \cflat \setminus 1_\cflat \} \right \rfloor.
$$
\end{IEEEproof}

\subsection{Generalized Singleton Bound for Polymatroids}

We can define the above notions analogously for polymatroids. In this section, we will  consider $\apart$-polymatroids and $\sapar$-polymatroids with information-symbol locality, 1-information-symbol locality, and all-symbol locality, as well as $\apoly$s with 1-information-symbol locality. 

The approach in several parts of the proof of the following theorem is similar to the one used for the corresponding bound  \cite{wang14b} for linear $(n,k,d,r,\delta=2,t)$-LRCs with systematic information-symbol locality.

\begin{thm} \label{th:bound_polymatroid} 
Let $\poly$ be an $\apar$-polymatroid with 1-information-symbol locality. Then 
$$
d \leq  n - \lceil k \rceil +1 - \left ( \left \lceil  \frac{(t ( \lceil k \rceil - 1)+1}{t(r-1)+1}\right \rceil - 1 \right ) (\delta - 1).
$$
\end{thm}

\begin{IEEEproof} 
Some parts of the proof are sketchy rather than rigorous. 
 Let $K$ be a 1-information set of $P$ with $\local$. By Theorem \ref{thm:parameters_via_Z}(iv), for each $x \in K$ there are $t$ repair sets $R_1(x),\ldots,R_t(x) \in \cset$ of $x$ with $(r,\delta)$-locality such that $i\neq j \Rightarrow R_i(x) \cap R_j(x) = \{x\}$. For $x \in K$ and $J \subseteq K $, let 
$$
R(x) = \bigcup_{i = 1}^t R_i(x) \com Z(x) = \cl(R(x)) \hand Z_J = \cl(\bigcup_{x \in J} Z(x).
$$
By Proposition \ref{proposition:recovering_sets}(iii) and (iv), 
$$
R(x) \com Z(x)\hand Z_J \in \cflat.
$$
We claim for any $x \in K$ and $i \in [t]$ that
$$
\begin{array}{cl}
(i) & \rho(R_i(x)) \leq |R_i(x)| - (\delta - 1) \leq r,\\
(ii) & \eta(R_i(x)) \geq \delta - 1,
\end{array}
$$
For statement (i), by Proposition \ref{prop:rank_1_poly}(i) and the definition of a repair set, we obtain that 
$$
\rho(R_i(x)) = \rho(Y) \leq |Y| = |R_i(x)| - (\delta - 1) \leq r 
$$
for any set $Y \subseteq R_i(x) \setminus \{x\}$ where $|Y| = |R_i(x)| - (\delta-1)$. Statement (ii) follows directly from statement (i).

We claim for any $x \in K$, $i \in [t]$ and $I \subseteq [t] \setminus \{i\}$ that 
$$
\begin{array}{cl}
(iii) & \rho(\bigcup_{l \in I} R_l(x)) \leq |I|(r-1) + 1,\\
(iv) & \eta(\bigcup_{l \in I} R_l(x)) \geq |I|(\delta-1).
\end{array}
$$
Statement (iii) follows from induction on $|I|$. The statement follows from statement (i) for $|I|=0$. Now, let $A = I \cup \{i\}$ and $y \in [t] \setminus A$, then by the induction assumption, and the axioms (R1) and (R3),
$$
\begin{array}{rcl}
\rho ( \bigcup_{l \in (A \cup \{y\} } R_l(x)) &
\leq & \rho(\bigcup_{l \in A} R_l(x) + \rho(R_y(x)) - \rho(x) \\
& \leq & |I|(r-1) + 1 + r - 1\\
& = & (|I|+1)(r-1) + 1.
\end{array}
$$
For statement (iv), by a similar argument as for statement (iii) above we have that 
$$
\rho(\bigcup_{l \in I} R_l(x)) \leq (\sum_{l \in I} \rho(R_l(x))) - 1. 
$$
Hence, (iv) follows from (ii) and (iii).

The property that $\rho(K) = k$ implies that $\rho(Z_K) = k$. Choose a subset $J = \{x_1,\ldots,x_j, x_{j+1}\} \subseteq K$ such that $\rho(Z_J) = k$ and $x_{i+1} \notin Z_{\{x_1,\ldots,x_i\}}$ for $1 \leq i \leq j$. For simplicity of notation, let $Z_{[i]}$ denote the cyclic flat $Z_{\{x_1,\ldots,x_i\}}$ for $1 \leq i \leq j+1$. 

By Statements (iii) and (iv) and Proposition \ref{prop:mixed_1_poly}(i) and (ii), we immediately obtain that
$$
\begin{array}{cl}
(iv) & \rho(Z(x_i)) = \rho(R(x_i)) \leq t(r-1) + 1,\\
(v) & \eta(Z(x_i) \geq \eta(R(x_i)) \geq t(\delta - 1),
\end{array}
$$
for $i \in [j+1]$. Hence, for $1 \leq i \leq j$,
$$
\begin{array}{rl}
(vi) & \rho(Z_{[i+1]}) - \rho(Z_{[i]}) \leq t(r-1) + 1,\\
(vii) & \eta(Z_{[i+1]}) - \eta(Z_{[i]}) \geq t(\delta-1).
\end{array}
$$
Statement (vi) is a consequence of (iv) and axiom (R3). 
Statement (vii) follows from the facts that $x_{i+1} \notin Z_{[i]}$.
Consequently, 
$$
\begin{array}{rl}
(viii) & \rho(Z_{[j]}) \leq j(t(r-1) + 1),\\
(ix) & \eta(Z_{[j]}) \geq jt(\delta-1).
\end{array}
$$
For $0 \leq l \leq t$, let
$$
Z_{j+1}^l = \cl(\bigcup_{i=1}^l R_i(x_{j+1}) ). 
$$
Now, let $s$ be the integer in $[t]$ such that 
$$
\rho(\cl(Z_{[j]} \cup Z_{j+1}^{s-1})) < k \hand  \rho(\cl(Z_{[j]} \cup Z_{j+1}^{s})) = k.
$$
The theorem now follows from similar arguments and enumerations as given in the proof of Theorem 1 i \cite{wang14}. 
\end{IEEEproof}

\begin{thm}  \label{corollary:bound 2}
Let $\poly$ be an $\apart$-polymatroid with information-symbol, 1-information-symbol, or all-symbol locality. Then 
$$
d \leq  n - \lceil k \rceil +1 - \left ( \left \lceil  \frac{(t ( \lceil k \rceil - 1)+1}{t(r-1)+1}\right \rceil - 1 \right ) (\delta - 1).
$$
\end{thm}

\begin{IEEEproof}
The proof of the results for information-symbol locality follow by similar argument as for Theorem \ref{th:bound_polymatroid}. The theorem now follows from the facts that every $\apart$-polymatroid with 1-information-symbol locality or all-symbol locality is an $\apart$-polymatroid with information-symbol locality.
\end{IEEEproof}

\begin{thm}  \label{corollary:bound 3}
Let $\poly$ be an $\sapar$-polymatroid with information-symbol, 1-information-symbol or all-symbol locality. Then 
$$
d \leq  n - \lceil k \rceil +1 - \left ( \left \lceil  \frac{(t ( \lceil k \rceil - 1)+1}{t(r-1)+1}\right \rceil - 1 \right ) (\delta - 1).
$$
\end{thm}

\begin{IEEEproof}
The proof of the results for information-symbol locality follow by similar argument as for Theorem \ref{th:bound_polymatroid}. The theorem now follows from the facts that every $\sapar$-polymatroid with 1-information-symbol locality or all-symbol locality is an $\sapar$-polymatroid with information-symbol locality.
\end{IEEEproof}

\subsection{Corollaries for LRCs}

From Theorems \ref{th:bound_polymatroid}, \ref{corollary:bound 2}, and \ref{corollary:bound 3} we immediately get the following  corollary for LRCs.

\begin{corollary} \label{corollary:bound_LRC_1}
Let $C$ be an $\apar$-LRC with 1-information-symbol locality. Then 
$$
d \leq  n - \lceil k \rceil +1 - \left ( \left \lceil  \frac{(t ( \lceil k \rceil - 1)+1}{t(r-1)+1}\right \rceil - 1 \right ) (\delta - 1).
$$
\end{corollary}

\begin{corollary} \label{corollary:bound_LRC_2}
Let $C$ be an $\apart$-LRC with 1-information-symbol, information-symbol or all-symbol locality. Then 
$$
d \leq  n - \lceil k \rceil +1 - \left ( \left \lceil  \frac{(t ( \lceil k \rceil - 1)+1}{t(r-1)+1}\right \rceil - 1 \right ) (\delta - 1).
$$
\end{corollary}

\begin{corollary} \label{corollary:bound_LRC_3}
Let $C$ be an $\sapar$-LRC with 1-information-symbol, information-symbol or all-symbol locality. Then 
$$
d \leq  n - \lceil k \rceil +1 - \left ( \left \lceil  \frac{(t ( \lceil k \rceil - 1)+1}{t(r-1)+1}\right \rceil - 1 \right ) (\delta - 1).
$$
\end{corollary}

One remark on the bounds given above is that, if all the parameters $(n,r,\delta,t)$ are fixed as well as the alphabet size $s$, then the bound for $d$ always increases when the number of codewords goes from $s^k$ to $s^k +1$. 
This, for example, implies that if there is a linear LRC that achieves some bound given above and there is a nonlinear LRC with the same parameters on $(n,r,\delta,t)$ but with a better rate then the nonlinear LRC will always have a smaller $d$ than the linear LRC.

Further, there are many polymatroids which cannot be realised as a polymatroid $P_C$ of any code $C \subseteq \alp^n$. For example the nonentropic polymatroids. Hence, the bounds given for polymatroids above are valid for many other types of polymatroids than just the $P_C$-polymatroids. The same is true for matroids, many of which are not representable by a linear code. In general, it is extremely hard to determine whether a given matroid is representable (over any field). It is conjectured, but to the best of the authors' knowledge not yet proven, that
$$
\lim_{n \rightarrow \infty} \frac{|\{\hbox{Representable matroids on $n$ elements}\}|}{|\{\hbox{Matroids on $n$ elements}\}|} = 0. 
$$

Moreover, there are many non-code objects that can be associated to matroids or polymatroids, \emph{e.g.}, graphs, hypergraphs, matchings, and designs. The bounds given above for polymatroids also give us results for all these additional objects. 

\section{Constructions of Perfect linear $\apar$-LRCs} \label{sec:construction}

Typically (Singleton-type) bound-achieving codes have been referred to as optimal. However, we rather choose to use the term perfect, since there might not always exist codes achieving the bound. However, in our interpretation, \emph{optimal} should always refer to the best option one can possibly have. Hence it feels wrong to us to say that no optimal code exists, even though there would be a code that almost achieves the bound and is known to be the best possible code. To this end, we give the following definition. 

\begin{definition}
We will call an $\apar$-polymatroid or $\apar$-LRC which achieves the bounds given above \emph{perfect}.
\end{definition}

In \cite{westerback15} a construction of linear LRCs is derived from matroid theory. This construction was used in  \cite{westerback15} to obtain linear $(n,k,d,r,\delta,t=1)$-LRCs with all-symbol locality that are perfect or near-perfect. We summarize the construction in the following. 

\begin{set_construction}[\cite{westerback15}] \label{set:construction}
Let $F_1,\ldots,F_m$ be subsets of a finite set $E$, $k$ a non-negative integer and $\rho:\{F_i\}_{i \in [m]} \rightarrow \mathbb{Z}$ a function such that 
\begin{equation} \label{eq:set_construction_conditions}
\begin{array}{rl}
(i) & 0 < \rho(F_i) < |F_i|,\\
(ii) & k \leq |F_[m]| - \sum_{i = 1}^m (\eta(F_i)),\\
(iii) &| F_{[m]\setminus \{i\}} \cap F_i| < \rho(F_i) \hbox{ for all } i \in [m], 
\end{array}
\end{equation}
where for every element $i \in [m]$ and subset $I \subseteq [m]$
$$
\begin{array}{rl}
(a) & \eta(F_i) = |F_i| - \rho(F_i),\\
(b) & F_I = \bigcup_{i \in I} F_i.
\end{array}
$$
Further, for every subset $I \subseteq [m]$, define
$$
\rho(F_I) = \min \{|F_I| - \sum_{i \in I} \eta(F_i), k\} \hand \rho(E) = k.
$$
\end{set_construction}

\begin{thm}[\cite{westerback15}] \label{theorem:matroid_construction}
Let $F_1,\ldots,F_m$ be subsets of a finite set $E$, $k$ a non-negative integer and $\rho:\{F_i\}_{i \in [m]} \rightarrow \mathbb{Z}$ a function such that the conditions (i)-(iii) in \eqref{eq:set_construction_conditions} are satisfied. Then the set-construction defines a matroid $M_\cflat = (\rho_\cflat,E)$ where
$$
\begin{array}{cl}
(i) & \cflat = \{F_I : I \subseteq [m] \com \rho(F_I) < k\} \cup E,\\
(ii) & \rho_\cflat(X) = \min \{\rho(F) + |X \setminus F| : F \in \cflat\},\\
(iii) & n = |E|,\\
(iv) & k = \rho(E),\\
(v) & d = n - k + 1 - \max \{\eta(F) : F \in \cflat \setminus \{E\}\},\\
(iv) & F_i \hbox{ is a repair set with}\\
      & \hbox{$(r = \rho(F_i), \delta = \eta(F_i)+1)$-locality for}\\
      & \hbox{every element in $F_i$},\\
(iv)  & \hbox{a subset $K \subseteq [n]$ is an information set of $M_\cflat \iff$} \\    & \hbox{$|K| = k$ and $|K \cap F| \leq \rho(F)$ for all $F \in \cflat$}.
\end{array} 
$$
\end{thm}

\begin{thm}[\cite{westerback15}]
Every matroid $M_\cflat$ given from Theorem \ref{theorem:matroid_construction} is in a class of matroids called gammoids.
\end{thm}

We say that a matroid is \emph{representable over a finite field} $\mathbb{F}_q$ if the matroid can be represented by a linear code over $\mathbb{F}_q$.

\begin{thm}[\cite{lindstrom73}] 
Every gammoid over a finite set $E$ is representable over every finite field of size greater then or equal to $2^{|E|}$. 
\end{thm}

We remark that $2^{|E|}$ is just an upper bound on the smallest field size of a linear code that can be used to represent a gammoid. It is possible that a gammoid may be represented by a linear code over a field with much less size than $2^{|E|}$.

The following theorem was derived in \cite{westerback15}, by use of the theorems above.

\begin{thm}[\cite{westerback15}] \label{theorem:M_Z_equals_M_C}
Every matroid $M_\cflat$ given in Theorem \ref{theorem:matroid_construction} is isomorphic to $M_C = (\rho_C,[n])$, for some linear code $C$ over a large enough field.
\end{thm}

Using Construction \ref{set:construction} and Theorem \ref{theorem:matroid_construction} we are now able to construct $\apar$-matroids $M_C$. To obtain the actual linear $\apar$-LRC associated to $M_C$ we can use \cite{lindstrom73} in which it is described how to derive a linear code associated to a gammoid. 

\begin{exam}{\it Construction of a perfect $\apar$-matroid $M_C$ for a linear code $C$.}

Let $E = [36]$, $k = 4$, 
$$
  \begin{array}{lcllcl}
    F_1 & = & \{1,5-8\}, & F_2 & = & \{1,9-12\},\\
    F_3 & = & \{2,13-16\}, & F_4 & = & \{2,17-20\},\\
    F_5 & = & \{3,21-24\}, & F_6 & = & \{3,25-28\},\\
    F_7 & = & \{4,29-32\}, & F_8 & = & \{4,33-36\},
   \end{array}
$$
and $\rho(F_i) = 3$ for $i \in [8]$. Then, by Theorem \ref{theorem:matroid_construction}, 
$$
\begin{array}{cl}
(i) & \cflat = \{\emptyset, F_1,\ldots,F_8,[36]\},\\
(ii) & K = \{1,2,3,4\} \hbox{ is an information set,}\\
(iii) & \hbox{for $i \in [8] \com F_i$ is a repair set with}\\
      & (r=3,\delta=3) \hbox{-locality for every element $x \in F_i$},\\
(iv) & d=36-4+1-2=31,\\
(v) & \hbox{$K$ has $(r=3,\delta=3,t=2)$-locality}.
\end{array}
$$
By Theorem \ref{theorem:M_Z_equals_M_C} and Corollary \ref{corollary:bound_LRC_1}, the construction above defines a perfect linear $(36,4,31,3,3,2)$-LRC with information-symbol locality since
$$
36-4+1 - (\left \lceil \left (\frac{2(4-1)+1}{2(3-1)+1}\right \rceil - 1 \right )(3-1)) = 31 = d.
$$
\end{exam}

\begin{thm}
If $n \geq k(t(r+ \delta-2) + 1))$, then there is a perfect linear $\apar$-LRC with information-symbol locality. 
\end{thm}

\begin{IEEEproof}
For a proof of the results above we use the same kind of construction given in the example above. A proof will appear in the journal version of this paper.
\end{IEEEproof}

\section*{Acknowledgments}
This work was partially  supported by the Academy of Finland grants  \#276031, \#282938, and \#283262, and by a grant
from Magnus Ehrnrooth Foundation, Finland. The support from the European Science Foundation under the ESF COST Action IC1104 is
also gratefully acknowledged.


\end{document}